\documentclass{JHEP3} 
\usepackage{amsmath}
\usepackage{epsfig}
\usepackage{amssymb,amsfonts}
\newcommand{\oao}[2]{{#1\atopwithdelims[]#2}}

\boldmath
\title{Quantization of heterotic strings in a G\"odel/Anti de Sitter 
spacetime and chronology protection\thanks{
Research partially supported by the EEC under the contracts
HPRN-CT-2000-00122, HPRN-CT-2000-00131.}}
\unboldmath
\author{Dan Isra\"el
\\
$\ $ \\
$\ $ \\
Laboratoire de Physique Th{\'e}orique
de l'Ecole Normale Sup{\'e}rieure\thanks{Unit{\'e} mixte  du
CNRS et de l'Ecole Normale Sup{\'e}rieure,
UMR 8549.} \\
$\;\;\;$24  rue Lhomond, 75231 Paris Cedex 05, France\\
$\ $ \\
E-mail:  \email{israel@lpt.ens.fr}
\\
}

\abstract{
We show that a G\"odel-like deformation  
of $AdS_3$ in heterotic string theory can be realized 
as an exact string background. 
Indeed this class of solutions is obtained 
as an exactly marginal deformation 
of the worldsheet 
conformal field theory describing the NS5/F1 heterotic
background. It can also be embedded in type II superstrings 
by Kaluza-Klein reduction. We compute the spectrum of this model as 
well as the genus one modular invariant partition function. We discuss 
the issue of closed timelike curves and the propagation of long strings. 
They destabilize completely the background, although we construct 
another exact string background that may describe the result of the 
condensation of these long strings. Closed timelike 
curves are avoided in that case.
}

\preprint{ 
LPTENS-03/32
\\hep-th/0310158}

\begin{document}

 
\section{Introduction}
The fate of closed timelike curves (CTC's) in general relativity is a long 
standing problem. There is in particular a conjecture by 
Hawking~\cite{Hawking:1991nk} \---~the ``Chronology Protection 
Conjecture''~\---  which states that, in quantum gravity, a causally safe 
background cannot develop closed timelike curves.  
One of the oldest known nontrivial spacetime 
with this pathology is the G\"odel universe~\cite{Godel:ga}. 
Spacetimes of this kind have been recently embedded in various supergravity 
theories~\cite{Gauntlett:2002nw}~\cite{Herdeiro:2002ft}~\cite{Boyda:2002ba}~\cite{Harmark:2003ud}, 
in the hope that string/M theory would cast a new light on the fate of 
closed timelike curves in these spacetimes. All these works are based on an analysis of the low
energy effective action and therefore do not take into account the
full quantum string effects. Nevertheless some interesting observations 
were made while considering extended BPS objects (branes and supertubes)
as probes in these spacetimes, which may destabilize the 
background~\cite{Drukker:2003sc}. 
Besides one type of G\"odel universe is T-dual to an exact type IIA pp-waves 
background. This opened the door for exact quantization of string 
probes~\cite{Russo:1994cv}~\cite{Russo:1995aj}~\cite{Harmark:2003ud}~\cite{Brace:2003st}, 
albeit only in the lighthcone gauge.

In this note we would like to point out that a one parameter 
class of spacetimes interpolating between $AdS_3$ and 
the G\"odel universe constructed in~\cite{Reboucas}
\--~and recently discussed in~\cite{Drukker:2003mg} in the context 
of general relativity~\--
can actually be embedded as an \emph{exact heterotic string background}. 
It turns out that it is obtained as an asymmetric deformation 
of the $SL(2,\mathbb{R})_L \times SL(2,\mathbb{R})_R$ conformal field theory 
(all the left-right symmetric deformations have been considered 
in a previous paper~\cite{Israel:2003ry}) and of the current 
algebra of the gauge group. If we want to consider type II 
superstrings \---~and in that case the background will 
be supersymmetric~\--- we can choose an internal compact $U(1)$ instead.  
The CFT description of this background allows us to compute 
the exact string spectrum and partition function. 
We can also trace back the different kind of states which appear 
in the spectrum in terms of  $SL(2,\mathbb{R})$ representations, 
and show that the no-ghost theorem for $SL(2,\mathbb{R})$ can 
be extended to the deformed theory.

Then we discuss long string states to 
see if something special happens due to the closed timelike 
curves around which they can wind. 
We will show explicitly that the long strings seems to destabilize 
the background, because their spectrum becomes highly tachyonic.
We propose a mechanism to solve the causality problem, by the 
condensation of a ring of fundamental strings, akin to the 
enhancon mechanism~\cite{Johnson:1999qt}.  
The endpoint of this condensation can also be realized as an exact 
conformal field theory, which does not contain instabilities as we shall  
see. This can be interpreted as a stringy chronology protection.    

The paper is organized as follows. In section~\ref{sigmamod}, 
we study the sigma model corresponding to the 
G\"odel/AdS$_3$ solution, i.e. the solution for the background 
fields. In section~\ref{spectrgodel} we compute the spectrum 
for all the string excitations in this background, and then 
the modular invariant partition function at genus one. 
In section~\ref{longstr} we consider 
the long string solutions, classically and at the level of 
the exact string background. 
We propose a scenario in section~\ref{gravdef} to solve the 
instability by condensing fundamental strings.  
Finally in section~\ref{concl} we summarize the main 
results of the paper and discuss holography. 
Appendix~\ref{slspec} is a short review of the spectrum and 
partition function for $AdS_3$; in appendix~\ref{coord} we give 
several coordinate systems on the G\"odel/AdS$_3$ spacetime. 
An outline of the proof of the no-ghost theorem is given in 
appendix~\ref{noghostads}.


\section{The G\"odel/AdS$_3$ solution}
\label{sigmamod}
\subsection{The worldsheet conformal field theory}
We start with the $AdS_3 \times S^3 \times K3$ supersymmetric 
background of heterotic string theory. It represents a configuration 
of wrapped NS5-branes and fundamental
strings~\cite{Antoniadis:1989mn}~\cite{Boonstra:1998yu}~\cite{Maldacena:1998bw}.
The part of the worldsheet action which is relevant to discuss 
the deformation comprises the $SL(2,\mathbb{R})_k$ \---~$AdS_3$~\--- part 
and the worldsheet right-moving current algebra :
\begin{eqnarray}
S & = & \frac{1}{4\pi} \int d^2 z \ \frac{k}{2} \left( \partial r \bar{\partial} r 
- \partial \alpha \bar{\partial} \alpha - 
\partial \beta \bar{\partial} \beta - 2 \cosh r \partial \alpha 
\bar{\partial} \beta \right) + 
\eta_{\mu \nu} \psi^{\mu} \bar{\partial} \psi^{\nu}  
+ \lambda^{I} \partial \lambda^{I} \nonumber \\
\end{eqnarray}
Here the $\psi^{\mu}$ are the worldsheet fermions of the 
holomorphic CFT (supersymmetric side), while the 
antiholomorphic fermions $\lambda^I$ are the internal 
fermions of the gauge sector, with currents: 
$\bar{J}^{IJ} = i \lambda^{I} (\bar{z}) \lambda^{J} (\bar{z})$.  
We would like to turn on a gauge field by switching on the
operator:
\begin{equation}
V (z,\bar{z} )= \left( J^3 + i \psi^+ \psi^- \right) \bar{J}^a \ ,
\label{margop}
\end{equation}
where
\begin{equation}
J^3 = k \left( \partial \beta + \cosh r \ \partial \alpha \right).
\end{equation}
Such a deformation is not possible around flat space because the 
corresponding operator:
$$V_{flat} = F_{ij}^a 
\left( \frac{1}{2} x^i \partial x^j + \psi^i \psi^j \right) \bar{J}^a
$$
is not marginal.
However a magnetic deformation of the NS5-branes background has been 
studied some years ago~\cite{Kiritsis:1995iu}. It was first 
introduced in the context of infrared regularization of  
superstrings~\cite{Kiritsis:1994ta}\footnote{Other magnetic 
backgrounds for the heterotic string have been constructed 
in \cite{Russo:1995aj}.}.\\
Our treatment of $AdS_3$ deformation will be close in
spirit. Switching on this gauge field in the bulk, and taking into 
account the back-reaction on the metric and the other fields, 
we obtain the solution of the sigma-model equations:
\begin{eqnarray}
\frac{4}{k} ds^2 & = & dr^2 - (1+2H^2 \cosh^2 r) d\alpha^2 - (1+2H^2)
d\beta^2 -2(1+2H^2)\cosh r \ d\alpha d\beta \nonumber \\
\nonumber \\
B & = & \frac{k}{4} \cosh r \ d\beta \wedge d\alpha \nonumber \\
A & = & -g \sqrt{k} H \left[ \cosh r \ d\alpha + d\beta \right]
\end{eqnarray} 
The metric can be also rewritten as a squashed $AdS_3$ geometry, as 
was first noticed in~\cite{Rooman:1998xf}: 
\begin{equation}
\label{squashed}
ds^2  = \frac{k}{4} \left[ dr^2 + \sinh^2 r d \alpha^2
- (1+2H^2) \left( d\beta + \cosh r \ d\alpha \right)^2 \right]
\end{equation}
The propagation of scalar fields in this spacetime has been 
recently addressed in~\cite{Drukker:2003mg}.
The original G\"odel spacetime is obtained for the special 
value of the deformation parameter: $$H_{Goedel} = \frac{1}{\sqrt{2}}.$$

The relation with $AdS_3$ in global coordinates becomes more obvious 
after the change of coordinates: 
$$\alpha = t + \phi \ , \ \ \beta = t-\phi \ , \ \ r = 2 \rho$$
Then we obtain the solution:
\begin{eqnarray}
\frac{1}{k} ds^2 & = & d\rho^2 
- \cosh^2 \rho  \ dt^2 
+ \sinh^2 \rho \ d\phi^2 
-2H^2 \left( \cosh^2 \rho \ dt + \sinh^2 \rho  \ d \phi \right)^2 
\nonumber \\
B & = & \frac{k}{2} \cosh 2\rho \ d\phi \wedge dt \nonumber \\
F & = & dA = g \sqrt{k} H \sinh 2\rho \ (dt + d\phi) \wedge d\rho
\end{eqnarray} 
By construction, the isometry group of this spacetime is 
$SL(2,\mathbb{R}) \times U(1)$.

\subsection{Causal structure}
This section is a short review of the results 
of~\cite{Rooman:1998xf}~\cite{Drukker:2003mg} that we give for 
completeness.  
The striking feature of the class of backgrounds~(\ref{squashed}) is that there are closed 
timelike curves spanned by the coordinate $\phi$ above the critical radius:
\begin{equation}
\label{CTCrad}
G_{\phi\phi} = k\sinh^2 \rho_c (1-2H^2 \sinh^2 \rho_c) = 0 
\ \to \ \rho_c = \mathrm{arcsinh} \frac{1}{\sqrt{2}H}
\end{equation}
These closed timelike curves are contractible, hence the spacetime is 
topologically trivial. 
Since the spacetime is homogeneous, they are closed timelike curves 
going through each point of the manifold.
Although translation in the coordinate $t$ corresponds to 
a globally defined timelike Killing vector, we cannot globally 
define a time function because the sections of the spacetime 
at constant $t$ are not spacelike hypersurfaces, 
due to the closed timelike curves. The Cauchy problem is not 
well-defined in this spacetime. 

The analysis of the geodesics in these backgrounds 
has been made in hyperbolic coordinates, see
appendix~\ref{coord}. The authors of~~\cite{Drukker:2003mg} 
observed that the null and timelike geodesics  projects 
onto full circles of arbitrary center. For the spacelike geodesics 
it is true only for those with $\rho > \rho_{c}$, i.e. those projecting 
onto closed timelike curves.
In $AdS_3$, it has been shown~\cite{Maldacena:2000hw} that the 
action of spectral flow on spacelike geodesics produces long strings 
that can reach the boundary of Anti de Sitter and wind around 
the $\phi$-coordinate~; in the G\"odel-deformed background, as we
will see, these long string states are still present, and can wind 
around the closed timelike curves, causing a potential instability.

\subsection{Conformal invariance}
We now have to show that the deformation of the conformal field theory 
is truly marginal to all orders in the deformation. We bosonize the 
right-moving gauge current used for the deformation as:
$\bar{J}^{a} = i\bar{\partial} Y$. 
We will work in bosonic strings for simplicity; in that case $\bar{J}^a$ has a 
left-moving partner. Then $Y(z,\bar{z})$ can be viewed as an internal coordinate, and 
the background will be obtained by a Kaluza-Klein reduction.
We write the relevant part of the
sigma-model for the deformed geometry in the Kaluza-Klein form as:
\begin{eqnarray}
\label{sighet}
S_{def} & = & \frac{1}{4\pi} \int d^2 z \ \frac{k}{2} \left\{ \partial r 
\bar{\partial} r 
- \partial \alpha \bar{\partial} \alpha - 
\partial \beta \bar{\partial} \beta - 2 \cosh r \partial \alpha 
\bar{\partial} \beta  \right. \nonumber \\
& & \left. -2H^2 \left[ \cosh^2 r \partial \alpha \bar{\partial} \alpha +
  \partial \beta \bar{\partial} \beta +  \cosh r \left(
\partial \alpha \bar{\partial} \beta +
\partial \beta \bar{\partial} \alpha 
 \right) \right]  \right\} \nonumber \\
& & +  
\left[ \partial Y - \sqrt{k} H \left(\cosh r \partial \alpha 
+ \partial \beta  \right) \right] 
\left[ \bar{\partial} Y - \sqrt{k} H \left(\cosh r \bar{\partial} \alpha 
+ \bar{\partial} \beta   \right) \right] \nonumber \\
\end{eqnarray}
To prove conformal invariance,  we note, as
in~\cite{Kiritsis:1995iu} that one may rewrite the action as: 
\begin{eqnarray}
\label{sigmarg}
S_{SL(2,R)} (r,\alpha ,\beta ) & + & \frac{1}{4\pi} 
\int d^2 z \ \partial Y \bar{\partial} Y -
  \frac{\sqrt{k} H}{2\pi} \int d^2 z \ \left( \partial \beta + \cosh r \ 
\partial \alpha \right) \bar{\partial} Y
  \nonumber \\
\label{ident}
& = & S_{SL(2,R)} \left( r,\alpha ,\beta + \frac{2H}{\sqrt{k}}\  Y \right) + 
\frac{1+2H^2}{4\pi} \int d^2 z \ \partial Y \bar{\partial} Y 
\end{eqnarray}
where we have dropped out a total derivative:
$$2\sqrt{k} H\int d^2 z \left( \partial Y \bar{\partial} \beta 
- \bar{\partial} Y \partial \beta \right);$$
this term will however give rise to topological contributions 
to the spectrum. So this background is an exactly conformal
invariant sigma-model, to all orders in $\alpha'$. 
It turns out that the last expression~(\ref{ident}) 
is exactly the same as the sigma-model of the deformed background, eqn.~(\ref{sighet}). 
In section~\ref{spectrgodel} we will compute explicitely  the spectrum 
for the heterotic background. 
We get also from eqn.~(\ref{sigmarg}) the 
modified expression for the $J^3$ current:
\begin{equation}
\label{defJ}
J^{3}_d  =  k \left( \partial \beta + \cosh r \partial \alpha \right)
+ 2H\sqrt{k} \partial Y \ ,
\end{equation}
and the $SL(2,\mathbb{R})_L$ symmetry is broken down to $U(1)$.  
Of course, the $SL(2,\mathbb{R })_R$ is preserved by the deformation.

\subsection{Target-space supersymmetry}
The supersymmetry variations for the gaugino, the dilatino and the
gravitino are:
\begin{eqnarray}
\delta \chi & = & F_{ab} \gamma^{ab} \ \epsilon \nonumber \\
\delta  \lambda & = & \left( \gamma^a \partial_a \phi - \frac{1}{6}
H_{abc} \gamma^{abc} \right) \epsilon \nonumber \\
\delta \Psi_{a} & = & \left[ \partial_a + \frac{1}{4} \left(
\omega_{a}^{bc} - H_{a}^{bc} \right) \gamma_{ab} \right] \epsilon
\end{eqnarray}
From the dilatino variation we obtain: 
\begin{equation}
\left[ \Gamma^3 \Gamma^4 \Gamma^5 - \frac{1}{\sqrt{1+2H^2}} 
 \Gamma^0 \Gamma^1 \Gamma^2 \right] \epsilon = 0
\end{equation}
where 0,1,2 are the tangent space indices for $AdS_3$ and 3,4,5 
those of $S^3$.
Thus, the supersymmetry is completely broken by the deformation. 
The analysis of the gaugino and the gravitino equations 
would lead to the same conclusion. However for the consistency of the 
string theory one still has to perform the GSO projection; the 
possible appearance of tachyonic modes will be addressed in 
section~\ref{longstr}.
If this solution is embedded in type II superstrings, as 
a Kaluza-Klein reduction \--~see eqn.~(\ref{sighet})~\-- the 
background is a supersymmetric solution of the corresponding 
Kaluza-Klein supergravity, as in~\cite{Caldarelli:2003a}. 
It will be shown in sect.~\ref{spectrgodel}.


\section{The string spectrum and one-loop partition function}
\label{spectrgodel}
The partition function for type IIB superstring theory on 
$AdS_3 \times S^3 \times K3$ was obtained in~\cite{Israel:2003ry}. 
The extension to heterotic case is straightforward, by using the 
heterotic map~\cite{Gepner:1987qi}. For concreteness, we consider 
an $E_8 \times E_8 \to E_8 \times E_7 \times SU(2) \times U(1)^4$  
compactification:
\begin{eqnarray}
Z (\tau) & = & 
{\mathrm{Im} \tau \over  \eta^2 \bar\eta^{14} }
Z_{SU(2)} Z_{SL(2,\mathbb{R})} {1\over 2} \sum_{h,g=0}^1
Z_{4,4} \oao{h}{g} \nonumber \\
&&\nonumber \\
&& \times {1 \over 2} \sum_{a,b=0}^1 (-)^{a+b}\
\vartheta^2 \oao{a}{b} \vartheta \oao{a+h}{b+g}
\vartheta \oao{a-h}{b-g} \nonumber \\
&&\nonumber \\
&& \times \ \bar{\Gamma}_{E_8} 
\times {1 \over 2} \sum_{\gamma, \delta=0}^1
\bar{\vartheta}
\oao{\gamma +h}{\delta +g} \bar{\vartheta} \oao{\gamma-h}{\delta-g}
\bar{\vartheta}^6 \oao{\gamma}{\delta} , \label{Zhet}
\end{eqnarray}
The twisted torus conformal blocks are: 
\begin{equation}
Z_{4,4}\oao{h}{g} =  \frac{\Gamma_{4,4}}{\eta^4
\bar{\eta}^4} \ \ \mathrm{for} \ \ h=g=0, \ \mathrm{and} \ Z_{4,4}\oao{h}{g}
= \frac{16\eta^2 \bar{\eta}^2}{\left\vert \vartheta\oao{1-h}{1-g}(0|
\tau)\right\vert^4}  \ \ 
\mathrm{otherwise}. \label{twbl}
\end{equation}
The $SU(2)_k$ bosonic conformal block is chosen to be the standard 
diagonal invariant:
\begin{equation}
Z_{SU(2)} = \sum_{l=0}^k \chi^{l}_k \bar{\chi}^{l}_k\ , 
\end{equation}
and the $SL(2,\mathbb{R})_{k+4}$ partition function is
given in appendix~\ref{slspec}.
\subsection{Exact string spectrum in the G\"odel /AdS spacetime}
Now we would like to perform the marginal deformation~(\ref{margop}) 
in this CFT heterotic background. As it is known generically for 
conformal field theories based on affine algebras, the deformation 
will act on the weight lattice of the subgroup along which the deformation 
is performed (see~\cite{Forste:2003km} for a recent discussion): 
\begin{itemize}
\item The Abelian \emph{elliptic subgroup} of the supersymmetric 
$SL(2,\mathbb{R})_{k+2}$, 
generated by $\mathcal{J}^3 = J^3 + i\psi^+ \psi^-$, for the left-movers, 
\item a $U(1)$ subgroup of the gauge group for 
the right-movers. 
\end{itemize}
The left and right weights of the relevant lattices are:
\begin{eqnarray}
\label{viras}
L_0 & = & - \frac{1}{k+4} \left( \tilde{m} + \frac{k+4}{2} w_+ \right)^2 
+ \frac{1}{2} \left( n + \frac{a}{2} \right)^2
\nonumber \\
& = & - \frac{1}{k+2} \left(  
 \tilde{m} + \frac{k+4}{2} w_+  + n + \frac{a}{2}  
\right)^2 + \frac{k+4}{2(k+2)} 
\left( n+\frac{a}{2} +w_+ + \frac{2\tilde{m}}{k+4} \right)^2 \\
\bar{L}_0 & = & 
- \frac{1}{k+4} \left( \bar{\tilde{m}} + \frac{k+4}{2} w_+ \right)^2
+ \frac{1}{2} \left( \bar{n} + \frac{\gamma}{2} \right)^2 
\end{eqnarray}
Here $n + a/2$ corresponds to the charges of the 
left fermionic current $i \psi^+ \psi^-$, 
and $a=0,1$ are respectively the NS and the R sector, as in the 
partition function. Also $\bar{n} + \gamma / 2$ corresponds to the 
weights lattice of the gauge current which is picked out by the deformation.
In the expression for $L_0$, eqn.~(\ref{viras}), 
we have explicitely factorized the 
supersymmetric weights of $SL(2,\mathbb{R})_{k+2}$. Then the deformation 
acts as a $O(2)$ rotation between these weights and the right 
weights of the $U(1)$ gauge field\footnote{it is of course the analytic 
continuation of the usual $O(1,1)$ transformation of toroidal lattices, 
but here the lattice of $J^3$ is timelike}, parameterized 
by the angle $\zeta$; we obtain then:
\begin{eqnarray}
\label{rotleft}
L_{0}^{def} & = & 
- \left[ \frac{\cos \zeta}{\sqrt{k+2}} \left(
\tilde{m} + \frac{k+4}{2} w_+  + n + \frac{a}{2}  \right)
+ \frac{\sin \zeta}{\sqrt{2}} 
\left(  \bar{n} + \frac{\gamma}{2} \right) 
\right]^2 \nonumber \\
& & + \frac{k+4}{2(k+2)} 
\left( n+\frac{a}{2} +w_+ + \frac{2\tilde{m}}{k+4} \right)^2 \\
\label{rotright}
\bar{L}_{0}^{def} & = & 
- \frac{1}{k+4} \left( \bar{\tilde{m}} + \frac{k+4}{2} w_+ \right)^2
\nonumber \\
& & + \left[  \frac{\cos \zeta}{\sqrt{2}} 
\left(  \bar{n} + \frac{\gamma}{2} \right) 
- \frac{\sin \zeta}{\sqrt{k+2}} \left(
\tilde{m} + \frac{k+4}{2} w_+  + n + \frac{a}{2}  \right)
\right]^2 
\end{eqnarray}
where the relation between this parameterization and the 
vev of the gauge field is:
\begin{equation}
\label{param}
\cos^2\zeta = \frac{1}{1+2H^2}
\end{equation}
Note that there is no maximal value for the gauge field, unlike 
the magnetic NS5-brane background~\cite{Kiritsis:1995iu}.\\
In the type II superstrings (Kaluza-Klein reduction), we replace 
the lattice from the right-moving heterotic gauge sector 
by the right lattice of a compact internal coordinate. 
In this case it is obvious that the deformation preserves 
8 supercharges coming from the right-moving fermionic sector, left 
untouched by the deformation.
\subsection{One-loop vacuum amplitude}
To compute the partition function for heterotic 
strings on G\"odel/AdS$_3 \times S^3 \times K3$ 
one simply has to combine 
the previous analysis of the spectrum with the remaining 
ingredients. Then the partition function of the deformed model reads 
(see appendix~\ref{slspec}):
\begin{eqnarray}
Z_{Goedel} (\tau) & = & 
{\mathrm{Im} \tau \over  \eta^3 \bar\eta^{15} }
Z_{SU(2)} {1\over 2} \sum_{h,g=0}^1
Z_{4,4} \oao{h}{g} \nonumber \\
&&\nonumber \\
&& \times {1 \over 2} \sum_{a,b=0}^1 (-)^{a+b}\
\vartheta \oao{a}{b} \vartheta \oao{a+h}{b+g}
\vartheta \oao{a-h}{b-g} \nonumber \\
&&\nonumber \\
&& \times \ \bar{\Gamma}_{E_8} 
\times {1 \over 2} \sum_{\gamma, \delta=0}^1
\bar{\vartheta}
\oao{\gamma +h}{\delta +g} \bar{\vartheta} \oao{\gamma-h}{\delta-g}
\bar{\vartheta}^5 \oao{\gamma}{\delta} \nonumber \\
&&\nonumber \\
&& \times \ \int d^2 t \ Z_{cigar} \oao{-t_1}{-t_2}
\sum_{N,W,n,\bar{n} \in \mathbb{Z}}
e^{i\pi \left(2Nt_2 + b(n+\frac{a}{2})  - \delta (\bar{n}+\frac{\gamma}{2})
\right)}\nonumber \\
&&\nonumber \\
&& \ \times \ q^{- \left[ \frac{\cos \zeta}{\sqrt{k+2}} \left(
\frac{N}{2} + \frac{k+4}{2} (W+t_1 )  + n + \frac{a}{2}  \right)
+ \frac{\sin \zeta}{\sqrt{2}} 
\left(  \bar{n} + \frac{\gamma}{2} \right) 
\right]^2 + \frac{k+4}{2(k+2)} 
\left( n+\frac{a}{2} +(W+t_1 ) + \frac{N}{k+4} \right)^2}\nonumber\\
&&\nonumber\\
&& \ \times \ \bar{q}^{- \frac{1}{k+4} \left( \frac{N}{2} - \frac{k+4}{2} (W+t_1) \right)^2
+ \left[  \frac{\cos \zeta}{\sqrt{2}} 
\left(  \bar{n} + \frac{\gamma}{2} \right) 
- \frac{\sin \zeta}{\sqrt{k+2}} 
\left( N + \frac{k+4}{2} (W+t_1)  + n + \frac{a}{2}  \right)
\right]^2
}
\end{eqnarray}
This partition function is modular invariant by construction since 
it is obtained as an $O(2,2,\mathbb{R})$
rotation of an even self-dual lattice. 
Due to the mixing between the fermionic characters and 
the lattice of $J^3$, it is not possible to use a Jacobi identity 
to prove target space supersymmetry; all the supersymmetry 
is broken in this background, as expected from  
the target space analysis.


\section{The fate of long strings}
\label{longstr}
The long strings, corresponding to the continuous representations 
of $SL(2,\mathbb{R})$ are the natural probes for the closed timelike 
curves since they can wind around the coordinate $\phi$ and 
become macroscopic. Such long strings in G\"odel universe 
T-dual to pp-waves have been considered recently 
in \cite{Brace:2003st}, both from the classical and quantum point of view in 
a lightcone gauge appropriate to this problem. 
No tachyons were found in that case, although the authors pointed out a 
possibility of unitarity violation.
Another analysis for M2-branes in M-theory G\"odel universes has been made 
in~\cite{Hikida:2003yd}.

\subsection{Classical analysis}
We would like to understand the behavior of simple long strings  
solutions.  We will consider classical solutions to the Nambu-Goto 
action:
\begin{equation}
S = -T_1 \int d\sigma d\tau \ \sqrt{G} + T_1 \int B_{[2]}
\end{equation}
\boldmath
\subsubsection{Long strings in AdS$_3$} 
\unboldmath
In the undeformed $AdS_3$ geometry, we parameterize a long 
string wrapping $w$ times the $\phi$ coordinate, 
without angular momentum, as:
\begin{equation}
\label{ansatz}
\left\{
\begin{array}{ccc}
t & = & w_+ \tau \\
\phi & = & w_+ \sigma \\
\rho & = & \rho( \tau )
\end{array}
\right.
\end{equation}
where we have used the static gauge for the time coordinate. 
So the classical Nambu-Goto action 
for a F1-string probe is: 
\begin{equation}
S =  - \pi k w_+ T_1 \int d\tau \left(  \sinh \rho \sqrt{w_{+}^2 
\cosh^2 \rho  - 
\dot{\rho}^2} - w_+ \cosh^2 \rho \right)
\end{equation}
with the equation of motion:
$$
-\frac{d}{d\tau} \frac{\dot{\rho}\sinh \rho}{\sqrt{w_{+}^2 \cosh^2 \rho -
    \dot{\rho}^2}}
= \frac{w_{+}^2 \cosh 2\rho -\dot{\rho}^2}{\sqrt{w_{+}^2 \cosh^2 \rho -
    \dot{\rho}^2}} \cosh \rho - w_+ \sinh 2\rho 
$$
We have a long string solution: $$\rho = w_+ |\tau |\ ;$$
so the long string reach the boundary of $AdS_3$ for $t = \pm \infty$.
Qualitatively this behavior appears  
because the attractive force 
due to the string tension and the repulsive force due to the NS-NS 
two-form cancel. This simple solution corresponds to the spectral flow 
of a spacelike geodesic.
More general solutions have being obtained using a group manifold approach 
in~\cite{Maldacena:2000hw}.  
\boldmath
\subsubsection{Long strings in AdS$_3$/G\"odel}
\unboldmath
Now we would like to examine long string solutions in the 
AdS/G\"odel space.  
The crucial feature is to check if 
such long strings can reach the critical radius and then 
wrap closed timelike curves.
We consider classical macroscopic strings which are 
uncharged under the gauge group for simplicity. 
As in the previous case, we are looking for 
solutions winding around $\phi$, without angular 
momentum, eqn.~(\ref{ansatz}).
Then the Nambu-Goto action is given by:
\begin{equation}
S = - \pi k w_+ T_1 \int d\tau \left(
\sinh \rho \sqrt{ (1+2H^2 )w_{+}^2 \cosh^2 \rho 
- (1-2H^2 \sinh^2 \rho )\dot{\rho}^2}
- w_+ \cosh^2 \rho \right)
\end{equation}
So we obtain the equation of motion for $\rho (\tau)$:
\begin{eqnarray*}
-\frac{d}{d\tau} \frac{(1-2H^2 \sinh^2 \rho )\sinh \rho \ \dot{\rho}
}{\sqrt{ (1+2H^2 )w_{+}^2 \cosh^2 \rho 
- (1-2H^2 \sinh^2 \rho )\dot{\rho}^2}} 
\ = \ \ \ \ \ \ \ \ \ \ \ \ \ \ \ \ \ \ \  \nonumber \\
 \frac{(1+2H^2) w_{+}^2 \cosh \rho \left( \cosh^2 \rho 
+ \sinh^2 \rho \right) + \left(
4H^2 \sinh^2 \rho -1 \right) \cosh \rho \dot{\rho}^2}
{\sqrt{ (1+2H^2 )w_{+}^2 \cosh^2 \rho 
- (1-2H^2 \sinh^2 \rho )\dot{\rho}^2}} \cosh \rho - 
 w_{+} \sinh 2 \rho \nonumber \\
\end{eqnarray*}
We still have also long string solutions with:
\begin{equation*}
\rho (\tau ) = \sqrt{1+2H^2} \ w_+ |\tau |
\end{equation*}
The radius of the worldsheet of these long strings 
grows linearly with time, until they eventually wrap 
the closed timelike curves.  
We conclude that the G\"odel deformation changes 
the on-shell values of the momenta, but otherwise leads to 
the same qualitative behavior of the long string solutions 
as in AdS$_3$: long strings can escape to infinity with a 
finite energy. 
\subsection{Unitarity of the physical spectrum} 
Naively, we would expect that the string spectrum would contain 
ghosts for long string solutions. In fact, the kinetic term 
for the field $\phi (z, \bar{z} )$ in the Polyakov action is:
$$\mathcal{L} \sim  k\sinh^2 r (1-2H^2 \sinh^2 r) \partial \phi 
\bar{\partial}{\phi},$$
so, for $r>r_c$ it seems that the sign of the kinetic term for the 
fluctuations of $\phi$ become negative. But the off-diagonal term 
between $\phi (z,\bar{z})$ and $t (z, \bar{z})$ makes the analysis 
more subtle; indeed, the determinant of the metric of the 
deformed spacetime is: 
$$-g = (1+2H^2) \left( \frac{k}{4} \right)^{3} \sinh^2 r,$$
which is always of the same sign.

To have a precise answer to this crucial issue, we will 
adapt the no-ghost theorem for $SL(2,\mathbb{R})$ , 
to the case of the G\"odel-deformed $SL(2,\mathbb{R})$. 
We would like to see if the modification of the 
structure of the zero modes changes the conclusion about 
the suppression of negative-norm states by the Virasoro 
conditions. A outline of the no-ghost theorem proof is given 
in appendix~\ref{noghostads}. The conclusion is: 
\emph{the spectrum doesn't contain negative-norm states}.

\subsection{Long strings spectrum}
Now we consider the exact string spectrum. We would like 
to see if the spectrum contains tachyons, i.e. on-shell states 
with an imaginary value for the energy\footnote{As we have seen 
there is no globally defined notion of energy in this spacetime. 
At any rate, we will take the same definition as in AdS$_3$ 
\---~the conjugate to the $t$ translations~\--- since the 
corresponding isometry is unbroken by the deformation.}:
\begin{equation}
E = \mathcal{J}_{0}^3 + \bar{J}_{0}^3  = 
\tilde{m} + \bar{\tilde{m}} + n + \frac{a}{2} + (k+2) w_+.
\end{equation} 
The mass-shell condition for the continuous representations 
reads ({\it ``other''} indicates the internal CFT): 
\begin{eqnarray}
\label{shellleft}
L_0  & = & 
\frac{s^2+1/4}{k+2} -w_+ \left( \tilde{m} + \tilde{n} + \frac{a}{2} \right)
 - \frac{k+2}{4} w_{+}^2 +  
\frac{1}{2} \left( \tilde{n} + \frac{a}{2} \right)^2  \nonumber \\
 & & +   \frac{\sin \zeta}{k+2}  \left\{ 
\left[ \left(\tilde{m}+\frac{k+2}{2}w_+ +\tilde{n}+\frac{a}{2}
\right)^2 -  \frac{k+2}{2} \left( \bar{n}+\frac{\gamma}{2} 
\right)^2 \right] \sin \zeta \right. \nonumber \\
& & \ \ - \left. \sqrt{2(k+2)} 
\left( \tilde{m}+\frac{k+2}{2}w_+ +\tilde{n}+\frac{a}{2}
\right) \left( \bar{n}+\frac{\gamma}{2} \right) \cos \zeta 
\vphantom{\left( \frac{1}{2} \right)^2} \right\}
 + N + h_{other} -\frac{1}{2}= 0
\nonumber \\
\\
\label{shellright}
\bar{L}_0 & = & \frac{s^2+1/4}{k+2} 
-w_+ \bar{\tilde{m}} - \frac{k+4}{4} w_{+}^2   
+\frac{1}{2}\left( \bar{n} + \frac{\gamma}{2} \right)^2
\nonumber \\
 & & +   \frac{\sin \zeta}{k+2}  \left\{ 
\left[ \left(\tilde{m}+\frac{k+2}{2}w_+ +\tilde{n}+\frac{a}{2}
\right)^2 -  \frac{k+2}{2} \left( \bar{n}+\frac{\gamma}{2} 
\right)^2 \right] \sin \zeta \right. \nonumber \\
& & \ \ - \left. \sqrt{2(k+2)} 
\left( \tilde{m}+\frac{k+2}{2}w_+ +\tilde{n}+\frac{a}{2}
\right) \left( \bar{n}+\frac{\gamma}{2} 
\right) \cos \zeta \vphantom{\left( \frac{1}{2} \right)^2}
\right\} + \bar{N} + \bar{h}_{other} -1= 0 \nonumber \\
\end{eqnarray}
In a supersymmetric WZW model, the spectral flow must act also 
on the fermions, so $n=\tilde{n} -w_+$ in the $w_+$-flowed sector 
(see~\cite{Pakman:2003cu}); it amounts to redefining the vacuum. 
We also have to act with the spectral flow in the gauge sector, 
in order to satisfy the matching condition $L_0 = \bar{L}_0$. 
At this point we have to decide if we act on the $U(1)$ along 
which we switch on a field strength
\--~\emph{choice (i)}~\-- or an other unbroken 
Cartan generator \--~\emph{choice (ii)}. 
Since we have constructed the deformation 
such that it preserves $L_0 - \bar{L}_0$, both 
choices are equally acceptable. However the spectrum is slightly
different.
In both cases, the matching condition will be satisfied, provided that: 
$$w_+ \left(\tilde{m} - \bar{\tilde{m}} \right) = w_+ \ell 
= h_{other} - \bar{h}_{other}, $$
where $\ell$ is the angular momentum.\\
We take for example the ``massless'' state:
\begin{itemize}
\item left~: $|s,p\rangle \otimes \psi^{other}_{-1/2} |0\rangle_{NS}$ 
\item right: $\bar{\alpha}^{other}_{-1} |s,\bar{p} \rangle \otimes  
|0\rangle_{NS}$  
\end{itemize}
Then we have, with the embedding of the spectral flow \emph{(ii):}
\begin{equation}
L^{(ii)}_0 = \frac{s^2+1/4+ \tilde{m}^2\sin^2 \zeta }{k+2}  -  
\frac{k+2}{4} w_{+}^2 \cos^2 \zeta 
 - w_+ \tilde{m} \cos^2 \zeta + \frac{1}{2} p^2
\end{equation} 
So we have the mass-shell condition for the state under consideration:
\begin{equation}
\label{massshell}
\frac{s^2 + 1/4}{k+2} + \frac{k+2}{4} w_{+}^2 - 
w_{+} \frac{E}{2} + \frac{\sin^2 \zeta}{4(k+2)} E^2 + \frac{1}{2} p^2 = 0  
\end{equation}
And 
\begin{equation}
E = \frac{k+2}{\sin^2 \zeta} \left(
w_{+} \pm \sqrt{w_{+}^2 \cos^2 \zeta - \frac{4\sin^2 \zeta}{k+2} 
\left( \frac{s^2 + 1/4}{k+2} + \frac{1}{2} p^2 \right)} \right).
\end{equation}
If for example we increase the mass of the state with some momentum 
$p$ in the compact internal CFT, the energy of the state will 
eventually become imaginary. 
This strange behavior is due to the quadratic term in the energy with 
a positive sign. The critical value is: 
$$\frac{s^{(ii)\ 2}+1/4}{k+2} + \frac{1}{2} p^2 = 
 \frac{k+2}{4} w_{+}^2 \mathrm{cotan}^2 \zeta.$$
For a more general state with oscillators, the same statement holds. An 
infinite number of massive long string states will become tachyonic 
when the G\"odel deformation is turned on (other tachyonic instabilities 
in magnetic fields have been discussed in~\cite{Russo:1995aj}); 
these instabilities are likely due to the wrapping of closed 
timelike curves.\\
For the other embedding of spectral flow \--~\emph{choice (i)}~\-- 
we have instead:
\begin{eqnarray}
L^{(i)}_0 & = &  \frac{s^2+1/4+ \tilde{m}^2\sin^2 \zeta }{k+2}  
   - \frac{1}{2}\left(
\sqrt{\frac{k+2}{2}} \cos \zeta + \sin \zeta \right)^2 w_{+}^2 
\nonumber \\
& & - \ \cos \zeta \left( \cos \zeta + \sqrt{\frac{2}{k+2}} \sin \zeta 
\right) w_+ \tilde{m} + \frac{1}{2} p^2
\end{eqnarray} 
and we find a correction to the critical value of momenta:
$$\frac{s^{(i)\ 2}+1/4}{k+2} + \frac{1}{2} p^2 = 
 \frac{k+2}{4} w_{+}^2 \left[\mathrm{cotan} \ \zeta
+ \sqrt{\frac{2}{k+2}} \right]^2 ,$$
but otherwise the conclusions are the same.
\paragraph{Behavior of short strings}$ $\\
Since the discrete representations correspond to strings trapped 
in the center of the spacetime (around $r = 0$), we expect that 
the corresponding spectrum is well-behaved. For the discrete 
representations, we have: $$is = \frac{1}{2} - \tilde{j},$$
and also a relation between the spin $\tilde{j}$ and the 
$\tilde{J}^3$ eigenvalue: $\tilde{m} = \tilde{j} + q$,
 with $q \in \mathbb{N}$, 
for the primaries of the lowest weight representations.  
The spin is also restricted for unitarity to 
\begin{equation}
\label{ubound}
1/2 < \tilde{j} < (k+3)/2
\end{equation}
We can solve the mass-shell condition for $\tilde{j}$: 
\begin{equation}
L^{\mathcal{D^+}}_0 = -\frac{\tilde{j} (\tilde{j} -1)}{k+2}  -  
\frac{k+2}{4} w_{+}^2 \cos^2 \zeta 
 - w_+ (\tilde{j}+q) \cos^2 \zeta + 
\frac{\sin^2 \zeta}{k+2} (\tilde{j} + q)^2 + h_{int} = 0, 
\end{equation} 
and find: 
\begin{eqnarray}
\tilde{j} & =& 
\frac{1}{2\cos^2 \zeta} \ - \ \frac{k+2}{2}  w_+ 
\ + \ q\ \tan^2 \zeta  \nonumber \\ 
& & +  \ 
\frac{1}{\cos^2 \zeta} \ \sqrt{ \frac{1}{4} 
+ (k+2) \left( N_{w_+} + h_{int} - \frac{w_+}{2} \right)
\cos^2 \zeta  
\ + \ q(q+1) \sin^2 \zeta}, 
\end{eqnarray}
where $N_{w_+} \equiv N -w_+ q$ is the level of the algebra after spectral flow. 
So we observe that the only effect of the G\"odel deformation will be that 
the range of values for the spin  
allowed by the unitary bound~(\ref{ubound}) will impose different 
constraints on the internal weights.
It would be interesting to relate this short string spectrum with 
the analysis of energy levels for a point particle in~\cite{Hikida:2003yd}
and~\cite{Drukker:2003mg}. It was observed that this setup can be mapped 
onto a Landau problem. 

\section{Can gravitational deformation cure the pathology ?}
\label{gravdef}
We have seen that the G\"odel deformation of AdS$_3$, 
supersymmetric or not, 
completely destabilizes the background. 
Then we expect some kind of closed tachyon condensation; 
the description of this process is out of reach at present, 
but we can try to find the background that would correspond to the 
endpoint of the tachyon decay. In other words we are looking 
for another point in the moduli space of worldsheet conformal 
field theories which could have this interpretation. 
\subsection{Switching on more background fields}
There is an additional deformation of the background which is 
exactly marginal, generated by the bilinear of the unbroken 
currents $$V_g \sim (J^3 + i\psi^+ \psi^- ) 
(\bar{J}^3 - i\bar{\psi}^+ \bar{\psi}^-).$$ We would like to check if 
such a deformation can eliminate the closed timelike curves. 
This deformation have been studied in~\cite{Israel:2003ry} around 
the $SL(2,\mathbb{R})$ point. Here we would like to perform this deformation 
\emph{in combination with G\"odel deformation}. This is possible 
since each deformation preserves the $U(1)_L \times U(1)_R$ 
symmetry generating the other one. 
The solution of the sigma-model equations  gives 
(see~\cite{Kiritsis:1995iu}):
\begin{eqnarray}
\label{gravimag}
\frac{4}{k} ds^2  & =  &  dr^2 
- \frac{(\lambda^2+1)^2 + (8H^2 \lambda^2 -
(\lambda^2 -1)^2) \cosh^2 r
}{\left( \lambda^2 + 1 + (\lambda^2-1)\cosh r
\right)^2} d\alpha^2 \nonumber \\
& & \ \ -  \frac{(\lambda^2+1)^2 + 8H^2 \lambda^2 -
(\lambda^2 -1)^2 \cosh^2 r}{\left( \lambda^2 + 1 + 
(\lambda^2-1)\cosh r \right)^2} d\beta^2 \nonumber \\
&& \ \ - 2 \frac{4\lambda^2 (1+2H^2)\cosh r - 
(\lambda^4 -1) \sinh^2 r}
{\left(\lambda^2 + 1 + (\lambda^2-1)\cosh r \right)^2} 
d\alpha d\beta\nonumber \\
B & = & \frac{k}{4}\frac{\lambda^2 -1 + (\lambda^2 + 1)\cosh r}
{\lambda^2 + 1 + (\lambda^2-1)\cosh r} d\beta \wedge d\alpha \nonumber \\
A & = & \frac{2g\sqrt{k} H \lambda}{\lambda^2 + 1 + (\lambda^2-1)\cosh r}
\left( d\beta + \cosh r d\alpha \right) \nonumber \\
e^{2\Phi} & = & \frac{\lambda \  e^{2\Phi_0}}{\lambda^2 + 1 
+ \left(\lambda^2 - 1\right) \cosh r} 
\end{eqnarray}
The model without $J^3  \bar{J}^3$ deformation corresponds 
to $\lambda = 1$. The scalar curvature of this manifold is given by:
\begin{equation}
\label{scalcurv}
R (H,\lambda ) = - \frac{8}{k} 
\frac{5\lambda^2 -1 -\lambda^4 - 2H^2 \lambda^2 +(1-\lambda^4 )\cosh r
}{(1+\lambda^2 + (\lambda^2 -1)\cosh r)^2}.
\end{equation}
We want to check if this spacetime has closed timelike curves. 
The $(\phi - \phi)$-component of the metric is:
\begin{equation}
G_{\phi \phi} = k\frac{
2\lambda^2 (1+2H^2)\cosh r + (1-\lambda^2 (1+2H^2))\cosh^2 r 
- (1+\lambda^2 (1+2H^2))}{(1+\lambda^2 + (\lambda^2 -1)\cosh r)^2}.
\end{equation}
We would like to know for which values of $\lambda$ the numerator 
is always positive. The conclusion is:
\begin{equation}
\label{causcond}
\mathrm{No \ CTC's} \leftrightarrow \lambda^2 \leqslant \frac{1}{1+2H^2}
\end{equation}
However, in this range, the background~(\ref{gravimag}) 
becomes singular. The scalar curvature, eqn.~(\ref{scalcurv}), 
blows up at: 
$$r_{sing} = \mathrm{arccosh}\  \frac{\lambda^2 +1}{1-\lambda^2}.$$ 
But the worldsheet conformal field theory is   
well-defined. This singularity corresponds to a ring of 
positive tension objects which are electrically charged 
under the NS-NS two form and charged under the gauge field. To have 
a more precise picture, we take for $\lambda$ the 
limiting value that avoids closed timelike curves: 
$\lambda^2 = 1/(1+2H^2)$. Then we find that the singularity is located at: 
\begin{equation}
r_{sing} = 2 \rho_{sing} = 2 \ \mathrm{arcsinh} \frac{1}{\sqrt{2}H}
\end{equation}
We observe that the locus of the annular singularity corresponds 
exactly to the critical radius above which closed timelike 
curves occurs, eqn.~(\ref{CTCrad}). Furthermore the long strings 
become tachyonic in the G\"odel/AdS$_3$ spacetime 
because they can wind around these closed timelike curves.
Therefore, it is very tempting to interpret this annular singularity 
as a ring of condensed fundamental strings, protecting the 
causally unsafe region, in the same spirit as the supertubes domain walls 
considered in~\cite{Drukker:2003sc}.
It would be interesting 
to make the connection with the D1/D5 black rings of~\cite{Elvang:2003mj}.
\subsection{Long string spectrum and chronology protection}$ $\\
Now we would like to know the expression of the 
long string spectrum as a function of the parameters of deformation 
$H$ and $\lambda$. By turning on the gravitational deformation 
$\lambda$, the mass-shell condition~(\ref{massshell}) becomes 
(see~\cite{Israel:2003ry}): 
\begin{equation}
L_0 = \frac{s^2 + 1/4}{k+2} + \frac{k+2}{4} w_{+}^2 - 
w_{+} \frac{E}{2} + \frac{1}{4(k+2)} \left( 1 - 
\frac{\cos^2 \zeta}{\lambda^2} \right) E^2 + h = 0  
\end{equation}
and the energy will be given by: 
\begin{equation}
E = (k+2) \left(1- \frac{\cos^2 \zeta}{\lambda^2} \right)^{-1} \left\{
w_{+} \pm \sqrt{w_{+}^2 \frac{\cos^2 \zeta}{\lambda^2} 
+ \frac{4}{k+2} \left( \frac{\cos^2 \zeta}{\lambda^2} -1 \right)    
\left( \frac{s^2 + 1/4}{k+2} + h \right)  } \right\}.
\end{equation}
so we find that there are no tachyonic behavior any longer 
in the spectrum, provided that $$\lambda^2 \leqslant \cos^2 \zeta .$$ 
Comparing with the relation~(\ref{param}), \emph{this is exactly the 
same as the condition~(\ref{causcond}) required to avoid closed 
timelike curves !} Therefore, we see that string theory 
offers a natural mechanism for chronology protection 
in G\"odel-like universes.

\section{Discussion and conclusions}
\label{concl}
We have studied a class of string backgrounds of the 
G\"odel type, which are homogeneous and contain closed timelike 
curves. We have shown that these solutions are in fact exact 
conformal field theories obtained by a truly marginal deformation 
of the $SL(2,\mathbb{R})$ CFT. In heterotic superstrings, this 
deformation switches on a non-trivial gauge field in the bulk, and breaks 
all the supersymmetry. Embedded in type II superstrings as a 
Kaluza-Klein reduction, this background preserves 8 supercharges 
coming from the right-movers, if we start with the $AdS_3 \times S^3 
\times K3$ background describing NS5 branes and fundamental strings. 
We were able to construct the exact string spectrum for this 
solution, as well as the modular invariant partition function. 
We showed that the G\"odel deformation preserves the unitarity 
of the physical spectrum, by proving a no-ghost theorem.  

We then discussed the issue of long strings, which can wrap the closed 
timelike curves, first by showing that such classical solutions still 
exists in the deformed background.
Then we studied the long string spectrum, expecting 
some pathology due to the closed timelike curves. We found that 
the spectrum of these long strings is highly tachyonic, for 
any nonzero value of the deformation parameter. We are aware that 
since there is no good definition of energy in this background, the 
notion of tachyon in this class of backgrounds is not completely 
well-defined. 
We expect that these long strings will destabilize the 
background. On the contrary, the spectrum of short strings is not 
pathologic, as we could infer because they are trapped in the center of 
the spacetime.

A possible resolution of the causality problem in this background 
has been proposed, by turning on another marginal deformation. It 
turned out that the closed timelike curves can be avoided, at the 
expense of going to a region of the moduli space where the background 
becomes singular. We observed that an annular singularity 
occurred precisely at the radius where the G\"odel/AdS space 
started developing closed timelike curves, which were the source 
of instability for the long string probes. We showed that 
the spectrum of the resulting background is well-behaved.  
Therefore, we have interpreted this curvature, NS-NS two form 
and gauge field singularity as the result of the condensation of a ring 
of fundamental heterotic strings, in the same spirit as the 
supertubes of~\cite{Drukker:2003sc}. 
The proper description of this 
process would require a second-quantized framework \--- 
closed string field theory. 
If this is indeed the proper interpretation, it would be the first example 
of a stringy chronology protection in an exact string theory framework. 

Since this background is continuously connected to $AdS_3 \times 
S^3 \times K3$ \---~the NS5/F1 background~\--- we can expect some 
holographic interpretation. This issue have already been discussed 
in~\cite{Boyda:2002ba}. The authors showed that if one applies the 
covariant prescription of Bousso~\cite{Bousso:1999cb}, on finds 
that the holographic screen, which is observer-dependent, encloses  
a \emph{the causally safe region}; 
they argued that it leads to an ``holographic protection'' 
of these spacetimes. The radius of this cylindrical screen was 
shown to be:
\begin{equation}
r_{screen} = 2 \ \mathrm{arcsinh} \frac{1}{2H}.
\end{equation}
This holographic screen is at finite distance, and the 
$H \to 0$ limit gives the usual holographic screen 
of $AdS_3$ at infinity. 
It is difficult to interpret the G\"odel deformation 
of AdS$_3$ in terms of the boundary theory of $AdS_3$,  
and to make the connection with the aforementioned proposal.
In fact, in AdS$_3$, the currents are non-normalizable operators and 
marginal current-current deformations corresponds to IR 
\emph{irrelevant} deformations of the dual CFT. Such an observation 
has been made in~\cite{Herdeiro:2002ft} for another class of 
G\"odel universe. 
Nevertheless, one might hope to understand the instability of 
the G\"odel/AdS spacetime due to long string states as some 
instability in the boundary theory, triggered by some 
irrelevant operator, and then have 
an holographic picture of the bulk tachyon condensation.
It have been shown~\cite{Herdeiro:2000ap} on a related 
example (BMPV black hole), that if we allow closed timelike 
curves in the bulk by increasing the rotation of the 
black hole, the states of the dual CFT will violate the 
unitarity bound. 
Since the spacetime we obtain after the condensation 
of the ring of fundamental strings seems stable, it 
would be really interesting to understand holography 
in this background, if any. We will address this 
problem in a future work.

\acknowledgments
I would like to thank C.~Bachas, Y.~Dolivet, M.~Petropoulos, 
B.~Pioline, J.~Troost and especially C.~Kounnas for very useful 
discussions and comments. I thank also M.~Berkooz and E.~Rabinovici 
for very interesting discussions about the first version of this paper.
I am grateful to C.~Kounnas, M.~Petropoulos 
and B.~Pioline for a careful reading of the manuscript.

\appendix
\boldmath
\section{The $SL(2,\mathbb{R})$ partition function}
\label{slspec}
\unboldmath
The computation of the one-loop vacuum amplitude has been 
done in~\cite{Israel:2003ry}, using a mixture of path integral 
and algebraic techniques. The result is, for 
the universal cover of $SL(2,\mathbb{R})$ at level $k+4$:
\begin{eqnarray}
\label{fctpartsl}
Z_{SL(2,\mathbb{R})} (\tau) 
& = & \frac{4 (k+4)\sqrt{k+2}}{\tau_{2}^{1/2}} 
\sum_{n,w,N,W}
\int_{0}^1 ds_1 ds_2 dt_1 dt_2
\frac{e^{\frac{2\pi}{\tau_2} 
(\mathrm{Im} ( s_1 \tau - s_2 ))^2}
}{\left|\vartheta_1 
\left( s_1 \tau - s_2  |\tau \right) \right|^2} \nonumber \\
\nonumber \\
& & \times \ 
e^{-\frac{(k+4)\pi}{\tau_2} 
\left| (s_1 - t_1 +w)\tau - (s_2 - t_2 + n) \right|^2}
e^{\frac{(k+4)\pi}{\tau_2} \left| (t_1 + W)\tau - (t_2 + N) \right|^2} 
\end{eqnarray}
The integration over the constraints $s_1$ and $s_2$ gives the 
spectra for all the flowed and unflowed representations 
(see~\cite{Israel:2003ry} for details). The spectrum of primary 
states gives:
\begin{itemize} 
\item \emph{Discrete representations} representing strings trapped 
in the center of $AdS_3$; they appear in the range $\frac{1}{2} < 
\tilde{j} <
\frac{k+3}{2}$. Their conformal weights are:
\begin{equation*}
L_0 = - \frac{\tilde{j}(\tilde{j}-1)}{k+2} + 
w_+ \left( -\tilde{m} - \frac{k+4}{4} w_+ 
\right) 
\end{equation*}
\item \emph{Continuous representations} with $\tilde{j}= -\frac{1}{2} + is$, 
$s \in \mathbb{R}_+$. They correspond to long strings that can reach 
the boundary of $AdS_3$. 
The spectral flow quantum number $w_+ = w + W$ represents their winding around 
the center near the boundary~\cite{Maldacena:2000hw}. The weights are:
\begin{equation*}
L_0 = \frac{s^2 +1/4}{k+2} + w_+ \left( -\tilde{m} - \frac{k+4}{4} w_+ \right) 
\end{equation*}
\end{itemize}
The relation between the unflowed and flowed eigenvalues follows from: 
$$\tilde{J}^{3}_n = J^{3}_n - \frac{k+4}{2} w_{+} \delta_{n,0},$$
and the spacetime energy is given by:
$$E = \tilde{m} + \bar{\tilde{m}} + (k+4) w_+ =  (k+4) (W + t_1).$$
For latter convenience, it is useful to rewrite the $SL(2,\mathbb{R})$ 
partition function as a twisted product of the coset 
 $SL(2,\mathbb{R})/U(1)$ (the ``cigar'' 2d Euclidean black hole)
and a timelike $U(1)$ : 
\begin{eqnarray}
Z_{SL(2,\mathbb{R})} & = &  \frac{1}{\eta \bar{\eta}}
\int d^2 t \ Z_{cigar} \oao{-t_1}{-t_2} \nonumber \\ & & \times \sum_{N,W}
q^{-\frac{1}{2} \left(\frac{N}{\sqrt{2(k+4)}} + 
\sqrt{\frac{k+4}{2}} (W+t_1) \right)^2}
\bar{q}^{-\frac{1}{2} \left(\frac{N}{\sqrt{2(k+4)}} -
\sqrt{\frac{k+4}{2}} (W+t_1) 
\right)^2}
e^{2i\pi N t_2},\nonumber\\ 
\end{eqnarray}
where (see~\cite{Hanany:2002ev}): 
\begin{eqnarray*}
 Z_{cigar} \oao{-t_1}{-t_2} & = & 
4 \sqrt{(k+2)(k+4)} \  
\sum_{m,w}
\int d^2 s \  
\frac{\eta \bar{\eta} \ e^{\frac{2\pi}{\tau_2} 
(\mathrm{Im} ( s_1 \tau - s_2 ))^2}}
{\left|\vartheta_1 
\left( s_1 \tau - s_2  |\tau \right) \right|^2}\\ 
&& \times 
\ e^{-\frac{(k+4)\pi}{\tau_2} 
\left| (s_1 - t_1 +w)\tau - (s_2 - t_2 + m) \right|^2}. 
\end{eqnarray*}

\section{Coordinate systems and geodesics}
\label{coord}
In cylindrical coordinates, the metric of the AdS/Go\"del form is
(see eqn.~(\ref{squashed}):
$$ds^2 = dr^2 + \sinh^2 r d\alpha^2 - \left( d\tau + 
4\Omega \sinh^2 (r/2) d\alpha \right)^2 \ \ 
\mathrm{with} \ \ 2 \Omega = \sqrt{1+2H^2}$$
One can go to Cartesian coordinates by the coordinate transformation:
\begin{equation*}
\left\{
\begin{array}{lcl}
e^{2x} & = & \cosh (2r) + \cos \left( \frac{\phi + \tau}{\Omega} \right) 
\sinh r \\
y e^{2x} & = & \frac{\Omega}{2} \sinh (2r)  
\sin \left( \frac{\phi + \tau}{\Omega} \right) \\
\tan \left( \frac{T+ (\phi - \tau )/2}{\Omega} \right)
& = & e^{-2r} \tan \left( \frac{\phi - \tau}{2\Omega} \right)
\end{array}
\right.
\end{equation*}
Then the metric reads:
$$ds^2 = dx^2 + \frac{1}{2} e^{2x} dy^2 
- \left( dT + \sqrt{2} \Omega e^x dy \right)^2.$$
The last useful coordinate system corresponds to the hyperbolic 
coordinates. It is obtained by:
$$x = - \ln X \ , \ \ y = \sqrt{2} Y$$
Then one obtains:
$$ds^2 = \frac{dX^2 + dY^2 - \left( X^2 dT + 2\Omega dY \right)^2}{X^2}
$$
In these coordinates, the geodesic equation gives~\cite{Drukker:2003mg}:
\begin{eqnarray*}
\dot{T} & = & (1+2H^2) E - \sqrt{1+2H^2} X p_x \\
\dot{X} & = & p_x X (Y - Y_0 ) \\
\dot{Y} & = & - \sqrt{1+2H^2} E X + p_x X^2
\end{eqnarray*}
where the energy $E$ is the charge associated to the Killing vector 
$\partial_T$, and $p_x$ the momentum associated to the Killing vector 
$\partial_x$.

\section{No-ghost theorem for string theory on G\"odel/AdS$_3$: 
sketch of a proof}
\label{noghostads}
Here we would like to outline the main steps of the no-ghost theorem 
for strings on $AdS_3$ and check whether it is still valid or not, in 
the case of the G\"odel deformation. 
Unitarity for a theory living in a spacetime containing 
closed timelike curves is maybe not a well-defined concept. 
Nevertheless, we would like to know if, as in $AdS_3$, the 
Virasoro constraints are able to remove all the 
negative-norm states from the worldsheet conformal field theory. 
We will follow~\cite{Petropoulos:1989fc}~\cite{Hwang:1991an}~\cite{Evans:1998qu}
and the appendix A of~\cite{Maldacena:2000hw}\footnote{It is also 
possible to use BRST methods, see~\cite{Pakman:2003kh} 
and~\cite{Asano:2003qb}.}. We consider string theory on 
(deformed) $SL(2,\mathbb{R})$ times a unitary ``internal'' CFT. 
The important observation is: writing the action for the deformed 
model as eqn.~(\ref{ident}), we see that, with 
a field redefinition, the actions of the deformed and the undeformed 
model are the same, \emph{except that the structure of the zero modes 
is different}. The mixing in the zero modes have been given in 
eqns.~(\ref{rotleft}) and~(\ref{rotright}). Note that the part 
of the CFT corresponding to the coset CFT 
$SL(2,\mathbb{R})_k / U(1)$ is left untouched.
Here we consider for simplicity the case of bosonic strings, and the 
G\"odel deformation is constructed as a Kaluza-Klein reduction 
along a coordinate $Y(z,\bar{z})$ compactified at the fermionic 
point:
$$Q_{y}^2 = \frac{1}{2} \left(p - \frac{q}{2}\right)^2 \ , \ \ 
\bar{Q}_{y}^2 = \frac{1}{2} \left(p + \frac{q}{2} \right)^2.$$
The extension to heterotic or type II superstrings is 
straightforward. 
Since only the zero mode structure differs from the 
$SL(2,\mathbb{R}) \times U(1) \times CFT_{internal}$, given a  
representation of the zero modes, the structure of the Verma modules 
and the affine representations are the same, provided 
that no new null vectors appear. For convenience 
we will include the spectral flow quantum number in the 
parafermionic theory\footnote{Usually the spectral 
flow is embedded in the $J^3 \bar{J}^3$ lattice, by expressing 
the spectrum in terms of $\tilde{m}$ rather than in terms of 
$m$. Nevertheless, our expression of the spectrum of the coset 
looks like in~\cite{Dijkgraaf:1991ba} and~\cite{Hanany:2002ev}.}, 
writing, for the undeformed $SL(2,\mathbb{R})_{k}$: 
\begin{equation}
\tilde{L}^{SL(2,R)}_0 = \tilde{L}^{SL(2,R)/U(1)}_0 - L^{(3)}_0 
\equiv \frac{c_2}{k-2} + \frac{1}{k} \left(m - \frac{k}{2} w_+ 
\right)^2 - \frac{m^2}{k}
\end{equation}  
where $c_2$ is the second Casimir of the group, and: 
$$L^{(3)}_n \equiv - \frac{1}{2k} \sum_m : J^{3}_m J^{3}_{n-m} :$$ 
The proof of the no-ghost theorem can be carried out in three steps. 
\paragraph{Step one:} we define $\mathcal{F}$ as the subspace 
of the Hilbert space $\mathcal{H}$
consisting in states $|f \rangle \in \mathcal{F}$
such that: $$L_n |f\rangle = J_{n}^3 |f \rangle = 0 \ 
\mathrm{for} \ n >0.$$
Then we have to show that the states of the form 
\begin{equation}
\label{vector}
L_{-n_1} \cdots L_{-n_N} J^{3}_{-m_1} \cdots 
J^{3}_{-m_M} |f \rangle \  \mathrm{with} \
n_1 \geq \cdots \geq n_N , \ m_1 \geq \cdots \geq m_M,
\end{equation} 
form a basis of $\mathcal{H}$. 
First, to prove that these states are linearly independent, 
we use an orthogonal 
decomposition of the Virasoro generators: 
$L_{n} = L_{n}^c - L_{n}^{(3)}$, 
and the fact that the states~(\ref{vector}) are in one to one 
correspondence with the similar states constructed with the $ L_{n}^c$'s.  
The Virasoro algebra corresponding to the $ L_{n}^c$'s has a central 
charge $c^c = 25$, and therefore the associated Verma modules will contain 
no null states, provided that the conformal weights are strictly positive.  
The conformal weights for the ``parafermionic'' theory  are, 
from eqns~(\ref{rotleft}) and~(\ref{rotright}):
\begin{eqnarray*}
L^{c}_0 & = & L_{0}^{SL(2,R)/U(1)} + \frac{1}{2} Q^2 + N_y 
+ h_{internal}, \\
\bar{L}^{c}_{0} & = &  \bar{L}_{0}^{SL(2,R)/U(1)} 
+ \frac{1}{2} \left( \bar{Q} \cos \zeta - \sqrt{\frac{2}{k}}\  m \ 
\sin \zeta \right)^2  
+ \bar{N}_y 
+ \bar{h}_{internal},
\end{eqnarray*}    
and it is known that the coset theory 
$SL(2,\mathbb{R})_k /U(1)$ is unitary and the weights are strictly positive, 
provided that the spin of the discrete representations is restricted to: $0<j<k/2$. 
So the states~(\ref{vector}) are linearly independent and form 
a basis of $\mathcal{H}$.\\
\paragraph{Step 2:} We have to show that a physical state can 
be chosen such that it can be written as~(\ref{vector}) with 
no $L_{-n}$.\\ 
Using the basis~(\ref{vector}), we write any state 
$|\psi \rangle$ of $\mathcal{H}$ as: 
$$|\psi \rangle = |sp \rangle + |\varphi \rangle,$$ 
where $| sp \rangle$ is a spurious state, i.e. 
with some $n_i$'s non zero, and $|\varphi \rangle$ 
has all its $n_i$'s being zero. 
Then we use the fact that, 
for a critical string background ($c = 26$), if 
we act on a on-shell (i.e. $L_0 = 1$) spurious state 
with $L_{n>0}$, we obtain again a spurious state. 
So we can map every physical state to a state 
$|\phi \rangle$ such that $L_{n>0} |\phi \rangle = 0$.
\paragraph{Step 3:} 
We have to show that if a physical state 
$| \varphi \rangle$ can be written as~(\ref{vector}) with 
no $L_{-n}$, then $J_{n>0}^3 |\phi \rangle = 0$.
First, this implies that
\begin{equation}
\label{primcond}
L^{(3)}_{n>0}  |\varphi \rangle = 0.
\end{equation}
We want to show that the states~(\ref{primcond}) are such that: 
$J_{n>0}^3 |\varphi \rangle = 0$.  
It will be true if there are no null states in the 
Virasoro descendents of $L^{(3)}$ for the left-movers, 
and $\bar{L}^{(3)}$ for the right-movers.  The weights 
of this $c=1$ conformal field theory are: 
\begin{eqnarray*}
L^{(3)}_0 & = & - \frac{1}{k} \left( m \ \cos \zeta + 
\sqrt{\frac{k}{2}} \  \sin \zeta \  \bar{Q} \right)^2 \\
\bar{L}^{(3)}_0 & = &  - \frac{1}{k} \bar{m}^2 
\end{eqnarray*}
So, this statement would be true 
if $L^{(3)}_0 \neq 0$ and $\bar{L}^{(3)}_0 \neq 0$ for all 
on-shell states. What remains to do is to 
examine separately the states with $J^{(3)}_0= 0$ 
or $\bar{J}^{(3)}_0 = 0$. 
\begin{itemize}
\item For the right algebra, the mass-shell 
condition~(\ref{shellright}) for $\bar{m}=0$ reads: 
$$
\bar{L}_0 = \frac{c_2}{k-2} + \frac{k}{4} w^2 + \bar{N}_{coset}
+ \frac{\cos \zeta}{2} \bar{Q}^2 + \bar{N}_3
+ \bar{h}_{internal} = 1
$$ 
where $\bar{N}_{coset}$ is the grade of the state with respect 
to the currents $\bar{J}^{\pm}$ and $\bar{N}_3$ is the grade 
w.r.t. the current $\bar{J}^3$. 
So, since $\bar{h}_{internal} >0$ there are 
no on-shell states obtained by the action 
of $\bar{J}_{-n}^3$ (i.e. with $\bar{N}_3 \neq 0$) 
for the continuous representations.     
It is also true for the discrete representations, provided 
that $0<\tilde{j}<k/2$, as in the undeformed $SL(2,\mathbb{R})$. 
\item For the left algebra, we are looking for states with 
\begin{equation}
\label{eigennull}
m \cos \zeta + \sqrt{k/2} \ \sin \zeta \ \bar{Q}  = 0.
\end{equation} 
We have the corresponding on-shell condition: 
$$
L_0 = \frac{c_2}{k-2} +  N_{coset} + \frac{1}{k}
\left(m -  \frac{k}{2} w_{+}  \right)^2 + N_3 
+ \frac{1}{2} Q^2 + N + h_{internal} = 1
$$
In a representation $\mathcal{D}_{\tilde{j}}^{+}$, we have 
$\tilde{m} = \tilde{j} + q$, $q \in \mathbb{Z}$, and, 
if $q$ is negative: 
\begin{equation}
\label{ineqgrade}
-q \leq N_{coset}.
\end{equation} 
So we have for the weights of the coset part: 
\begin{eqnarray*}
L_{0}^{coset} & = & \frac{-\tilde{j} (\tilde{j} -1 )}{k-2} 
+ \frac{\tilde{m}^2}{k} + N_coset \\
& = & \frac{q^2}{k} + 
\frac{\tilde{j}}{k-2} \left[ 1 - \frac{2\tilde{j}}{k} \right] 
+ \frac{2}{k} \tilde{j} q + N_{coset}
\end{eqnarray*}
If $\tilde{j}$ is taken in the range $0<\tilde{j}<k/2$, then with the 
inequality~(\ref{ineqgrade}) we obtain that 
$L_{0}^{coset} > 0$. So there are no on-shell states 
with $N_3 \geq 1$.  
\end{itemize}
This completes the proof of the no-ghost theorem.


\end{document}